\documentclass{ws-procs975x65}
\usepackage{amssymb,graphicx}

\def\be{\begin{equation}}
\def\ee{\end{equation}}
\def\bea{\begin{eqnarray}}
\def\eea{\end{eqnarray}}
\def\grad{\mathbf{\nabla}}

\def\v{\mathbf{v}}
\def\x{\mathbf{x}}

\def\U{\mathbf{U}}

\def\J{{\cal J}}

\def\P{{\cal P}}
\def\E{{\cal E}}
\def\L{{\cal L}}
\def\T{{\cal T}}
\def\R{{\cal R}}

\def\pd{\partial}
\def\d{\mathrm{d}}

\def\exp{\mathrm{exp}}

\def\etal{\textit{et.al.} \,}
\def\ie{\textit{i.e.} \, }

\begin{document}

\title{\bf \large Fluid QCD approach for quark-gluon plasma in stellar structure}

\author{T.P. Djun$^{a}$ and L.T. Handoko$^{a,b}$}

\address{$^{a)}$Group for Theoretical and Computational Physics, Research
Center for Physics, Indonesian Institute of Sciences, Kompleks
Puspiptek Serpong, Tangerang 15310, Indonesia}

\address{$^{b)}$Department of Physics, University of Indonesia,
Kampus UI Depok, Depok 16424, Indonesia }

\begin{abstract}
The quark-gluon plasma in stellar structure is investigated using the fluid-like QCD approach. The classical energy momentum tensor relevant for high energy and hot plasma having the nature of fluid bulk of gluon sea is calculated within the model. The transition of gluon field from point particle field inside stable hadrons to relativistic fluid field in hot plasma and vice versa is briefly discussed. The results are applied to construct the equation of state using the Tolman--Oppenheimer--Volkoff equation to describe the hot plasma dominated stellar structure.
\end{abstract}


\bodymatter

\section{Introduction}
\label{sec:intro}

Recent experiments in the last decades on relativistic nuclear collisions shed light on the phenomena of hot plasma formed by dense quarks and gluons. Those experiments suggest that the quark gluon matter behaves more like a deconfined quark-gluon plasma (QGP) liquid \cite{star,phenix}. A comprehensive review on this matter is given by E. Shuryak \cite{shuryak}. 

This fact immediately encourages some models based on the (relativistic) hydrodynamic approaches. In particular, dissipative ideal hydrodynamics has been used to fit some experimental data at high energy heavy ion program at the Relativistic Heavy Ion Collider (RHIC) \cite{rhic}. The successful fit requires the models to take into account very small value of the ratio shear viscosity over entropy \cite{teaney,huovinen,kolb,kolb2,hirano,baier}. However the puzzle must still be confirmed by the next coming experiments at the Large Hadron Collider (LHC) \cite{lhc}.

Since QGP is containing many quark-anti-quarks and gluons, it is considerable to treat it using the well-established quantum chromodynamics (QCD). In pure QCD, QGP is described as a quark soup before hadronization which is a phase of QCD, and exists at extremely high temperature  and/or density. It is argued that this phase consists of almost free quarks and gluons. Therefore, the phase transition from the deconfined QGP to the hadronic matters or vice versa gets particular interest in this approach. It unfortunately turns out to the many body problems with large color charge which cannot be calculated analytically using perturbation. As a result, the main theoretical tools to explore the QGP within QCD is lattice gauge theory. The lattice calculation predicts that the phase transition occurs at approximately 175 MeV \cite{gottlieb,petreczky}. 

On the other hand concerning that the QGP is a strongly interacting elementary particle system which should be governed by strong interaction, while it also dissolves into an almost perfect dense fluid of quarks and gluons \cite{zajc}, it is plausible to describe it as a fluid system. In this sense, there are approaches based on unifying or  hybridizing the charge field with flow field \cite{heinz, holm, choquet, blaizot, bistrovic, mahajan1, manuel, mahajan2}. Recently, some works have constructed the models in a lagrangian with certain non-Abelian gauge symmetry to the matter inside the fluid \cite{marmanis,sulaiman}. 

In this paper, the energy momentum tensor of QGP within the recent fluid QCD model\cite{sulaiman} is investigated. Further, the equation of state relevant for hot plasma dominated stellar structure is constructed using the so-called Tolman--Oppenheimer--Volkoff (TOV) equation.

The paper is organized as follows. First we briefly introduce the underlying model of gauge invariant fluid lagrangian and discuss the relevant physical scale and region within the model. Then, the energy momentum tensor in the model is derived and investigated. Subsequently it is followed with relevant equation of state in a particular geometry using TOV equation \cite{tolman,ov}. Finally, the paper is ended with a summary.

\section{The model}
\label{model}

Let us adopt the model developed by Sulaiman \etal\cite{sulaiman}. It describes the QGP as a strongly interacting gluon sea with the matters of quarks and anti-quarks inside. The model deploys the conventional QCD lagrangian with SU(3) color gauge symmetry, that is,
\be 
  \L = i \bar{Q} \gamma^{\mu} \pd_\mu Q - m_Q \bar{Q} Q -
  \frac{1}{4} S^a_{\mu\nu} {S^a}^{\mu\nu} + g_s J^a_\mu {U^a}^\mu \, . 
  \label{eq:l}
\ee
Here $Q$ and $U_\mu$ represent the quark (color) triplet and gauge vector field. $g_s$ is the strong coupling constant, $J^a_\mu = \bar{Q} T^{a} \gamma_\mu Q$ and $T^a$'s belong to the SU(3) Gell-Mann matrices. The strength tensor is $S^a_{\mu\nu} = \pd_\mu U^a_\nu - \pd_\nu U^a_\mu + g_s f^{abc} U^b_\mu U^c_\nu$ with $f^{abc}$ is the structure constant of SU(3) group respectively. It should be noted that the quarks and anti-quarks feel the electromagnetic force due to the U(1) field $A_\mu$, but the size is suppressed by a factor of $e/g = \sqrt{{\alpha}/{\alpha_s}} \sim O(10^{-1})$. 

Following the original model\cite{sulaiman}, the gluon fluid is put to have a particular form in term of relativistic velocity as,\footnote{This form was first proposed in the early work of Sulaiman \etal in 2005 available in arXiv:physics/0508219. The form was then adopted in the work of Bambah \etal in arXiv:hep-th/0605204, but the citation to the original work disappeared in their published version \cite{mahajan2}.}
\be
	U_\mu^a = (U^a_0, \U^a) \equiv u^a_\mu \, \phi \; ,
	\label{eq:u}
\ee
with $u_\mu^a \equiv \gamma_{\v^a} (1, \v^a)$ and $\gamma_{\v^a} =  {(1 - |\v^a|^2)}^{{-1}/2}$. $\phi$ is a dimension one scalar field to keep correct dimension and should represent the field  distribution. It is argued that taking this form leads to the equation of motion (EOM) for a single gluon field as follow\cite{sulaiman},
\be
	\frac{\pd}{\pd t} \left( \gamma_{\v^a} \v^a \phi \right) + \grad \left( \gamma_{\v^a} \phi \right) = -g_s \oint \d \x \, \left( \J^a_0 + F^a_0 \right) \, ,
	\label{eq:ree}
\ee
where $\J^a_\mu$ is the covariant current of gluon field, and $F^a_\mu$ is an auxiliary function which can be found in the original paper\cite{sulaiman}. It has been concluded that Eq. (\ref{eq:ree}) should be a general relativistic fluid equation, since at the non-relativistic limit Eq. (\ref{eq:ree}) coincides to the classical Euler equation. 

More precisely, Eq. (\ref{eq:ree}) provides a clue that a single gluonic field $U^a_\mu$ may behave as a fluid at certain scale, beside its point particle properties with a polarization vector $\epsilon_\mu$ in the form of  $U^a_\mu = \epsilon^a_\mu \, \phi$. One can consider that there is a kind of ``phase transition'',
\be
  \underbrace{\mathrm{hadronic \; state}}_{\displaystyle \epsilon^a_\mu} \longleftrightarrow \underbrace{\mathrm{QGP \; state}}_{\displaystyle u^a_\mu} \; .
\ee
As the gluon field behaves as a point particle, it is in a stable hadronic state and is characterized by its polarization vector. On the other hand in the pre-hadronic state (before hadronization) like hot QGP, the gluon field  behaves as a highly energized flow particle and the properties are dominated by its relativistic velocity. 

One should also recall that the wave function $U_\mu$ for a free particle satisfies $\left[ g^{\nu\mu} ( \pd^2 + m_U^2) - \pd^\nu \pd^\mu \right] U_\mu = 0$ with a solution $U_\mu \sim \epsilon_\mu \, \mathrm{exp} (-i p_\nu x^\nu)$ where $p_\nu$ is the 4-momentum. For a massive vector particle, \ie $m_U \neq 0$, we have no choice but to take $\pd^\mu U_\mu = 0$. It is not a gauge condition like the case of massless particle. This then demands $p^\mu \epsilon_\mu = 0$. Therefore the number of independent polarization vectors is reduced from four to three in a covariant fashion. In contrast with this, in the case of massless bosonic particles like gluon there are only two degrees of freedom remain. Therefore, one should keep in mind that in the present model the spatial velocity has only 2 degrees of freedom, that means one component must be described by another two vector components. Fortunately, in real applications in cosmology or compact star, this requirement is satisfied by the assumption that the system under consideration is isotropic.

From now, throughout the paper let us focus only on the gluon sea of plasma. This means one should consider only the related gluonic terms in Eq. (\ref{eq:l}), 
\be 
  \L_g = -\frac{1}{4} S^a_{\mu\nu} {S^a}^{\mu\nu} + g_s J^a_\mu {U^a}^\mu \, . 
  \label{eq:lg}
\ee

\section{Energy momentum tensor}
\label{emt}

Now we are ready to proceed with deriving the energy momentum tensor within the model. It should be pointed out that once the hot (high energy) QGP state is achieved, the system is assumed to be predominated by the classical motion rather than the quantum effects.

Therefore the total action of matter for non-gravitational fields in a general geometry of space-time $\R$ is $S_g = \int_\R \d^4x \, \sqrt{-g} \, \L_g$, where $g$ is the determinant of metric $g_{\mu\nu}$. It is well-known that the variation of $S_g$ in the metric is given by $\delta S_g = -\frac{1}{2} \int_\R \d^4x \, \sqrt{-g} \, \T_{\mu\nu} \, \delta g^{\mu\nu}$. Since the energy momentum tensor density is, 
\be
  \T_{\mu\nu} = \frac{2}{\sqrt{-g}} \frac{\delta \L_g}{\delta g^{\mu\nu}} \, ,
\ee
one obtains,
\be
  \T_{\mu\nu} = S^a_{\mu\rho} {S^a}_\nu^\rho - g_{\mu\nu} \L_g + 2 g_s J^a_\mu {U^a}_\nu \, . 
  \label{eq:t}
\ee
It is clear that Eq. (\ref{eq:t}) is symmetric as expected to fulfill the Einstein gravitational EOM. The total energy momentum tensor $T_{\mu\nu}$ is given by integrating out Eq. (\ref{eq:t}) in term of total volume in the space-time under consideration. $T_{\mu\nu}$ is a result of bulk of gluons flow in the system.

Furthermore, in a general space-time coordinates, the components of energy momentum tensor determine the total energy density ($T_{00}$), the heat conduction ($T_{0i,i0}$), the isotropic pressure ($T_{ii}$) and the viscous stresses ($T_{ij}$ with $i \neq j$) of the gluonic plasma. Of course, in this case the derivative $\pd_\mu$ inside the strength tensor $S_{\mu\nu}$ should be replaced by the covariant one, $\grad_\mu$. Also, the energy momentum tensor satisfies the conservation condition, $\grad_\mu \, \T^{\mu\nu} = 0$. Nevertheless, one can trivially conclude that the model induces non-zero viscosity since generally $T_{ij} \neq 0$ for $i \neq j$. From the experimental clues, however the size should be small such that it is always treated perturbatively in most hydrodynamics models\cite{teaney,huovinen,kolb,kolb2,hirano,baier}.

Before going further to apply these results, one should determine the quark current $J^a_\mu$ in Eq. (\ref{eq:t}). This can be simply calculated by considering the EOM (Dirac equation) of a single colored quark ($q$) or anti-quark ($\bar{q}$) with 4-momentum $p_\mu$. Since the solution of the EOM is $q(p,x) = u(p) \, \exp(-i p \cdot x)$, one immediately gets $\bar{u} \gamma_\mu u = 4 p_\mu$. Assuming that all colored quarks / anti-quarks have the same momenta and  the velocity of gluons are homogeneity, approximately $J^a_\mu {U^a}^\mu \propto 4 p_\mu U^\mu = 4 m_Q \phi$ since $u_\mu u^\mu = u^2 = 1$.

\section{Equation of state for stellar structure}
\label{eos}

This section is dedicated to provide an example on the applications of the present model to describe the stellar interior, in particular the compact stars which are still dominated by hot plasma before transforming itself into neutron star.

The stellar structure is commonly described as a static spherically symmetric space-time represented by Schwarzschild geometry. This means one deals with the relativistic gravitational equations for the interior of spherically symmetric plasma distribution. In the region under consideration the presence of the flow gluonic fields induces non-zero energy momentum tensor which is making up the star. This is the phase before the neutron star is getting mature. Starting from the stellar nebula made of hot plasma which is gradually getting colder as the hadronization occurs from the colder surface, while the inner core is still in pure hot QGP state.

As a consequence of the diagonal metric of Schwarzschild space-time, the model falls back to the perfect fluid without viscosity and heat conduction, \ie $T_{0i} = T_{ij} = 0$ for $i \neq j$. Also, since the plasma distribution should be spherically isotropic, it is considerable to put $v_1 = v_2 = v_3 = v$ as constant for all colored gluons. This assumption is consistent with the degree of freedom counting discussed in the preceding section. Moreover, the vanishing off-diagonal components of the Ricci tensor, $R_{i0}$, actually forces the spatial 3-velocity of the fluid must vanish everywhere. Hence particular assumption for $v_i$ is indeed not necessary. However, the gluon distribution still depends on the radius length, $\phi = \phi(r)$. 

For the sake of simplicity one can put homogeneous gluon fields for all color states, \ie $U_\mu^a = U_\mu$ for all $a = 1, \cdots, 8$. This yields,
\bea
  \T_{\mu\nu} & = & \left[ 8 \, g_s \, f_Q \, m_Q \, \phi(r) + g_s^2 \, f_g^2 \, \phi(r)^4 \right] u_\mu u_\nu  
  \nonumber\\
  && - \left[ 4 \, g_s \, f_Q \, m_Q \, \phi(r) - \frac{1}{4} g_s^2 \, f_g^2 \, \phi(r)^4 \right] g_{\mu\nu} \, ,
  \label{eq:tf}
\eea
where $f_g$ is the factor of summed colored gluon states from the structure constant $f^{abc}$, while $f_Q$ is the factor of summed colored quark states from $J^a_\mu {U^a}^\mu$. Remind that the energy momentum tensor for perfect fluid takes the form,
\be
  \T_{\mu\nu} = \left( \E + \P \right) u_\mu u_\nu  - \P \, g_{\mu\nu} \, .
  \label{eq:tpf}
\ee
Here $\E$ and $\P$ denote the density and isotropic pressure for single fluid field, each is related to the total density and pressure of the system through $\rho = \oint \d^4 x \, \E$ and $P = \oint \d^4 x \, \P$ respectively.
Obviously, from Eqs. (\ref{eq:tf}) and (\ref{eq:tpf}) one can obtain the density and pressure in the model as follows,
\bea
  P(r) & = & \int^\beta_0 \d t \int  \d V \left[ 4 \, g_s \, f_Q \, m_Q \, \phi(r) - \frac{1}{4} g_s^2 \, f_g^2 \, \phi(r)^4 \right] 
  \nonumber\\
  & = & \frac{4 \, g_s \, f_Q \, m_Q}{T} \int  \d V \left[ 1 - \frac{g_s \, f_g^2}{16 \, f_Q \, m_Q}  \phi(r)^3 \right] \phi(r) \, , 
  \label{eq:p}\\
  \rho(r) & = & \int^\beta_0 \d t \int  \d V \left[ 4 \, g_s \, f_Q \, m_Q \, \phi(r) + \frac{5}{4} g_s^2 \, f_g^2 \, \phi(r)^4 \right] 
  \nonumber\\
  & = & \frac{4 \, g_s \, f_Q \, m_Q}{T} \int  \d V \left[ 1 + \frac{5 \, g_s \, f_g^2}{16 \, f_Q \, m_Q}  \phi(r)^3 \right] \phi(r) \, , 
  \label{eq:e}
\eea
at a finite temperature $\beta = 1/{T}$ in a 3-dimensional spatial volume $V$. 

The proper spatial volume element for Schwarzschild geometry is $\d V = \sqrt{B(r)} r^2 \, \sin\theta \, \d r \, \d \theta \, \d \varphi$ with radius $r$ and two angles $\theta$ and $\varphi$ in spherical coordinates. The solution for $B(r)$ is given by,
\be
  B(r) = \left[ 1 - \frac{2 G m(r)}{r} \right]^{-1} \, ,
  \label{eq:b}
\ee
and $m(r) = 4 \pi \int^r_0 \d\bar{r} \rho(\bar{r}) \, \bar{r}^2$ is the 'bare mass'. This generates the proper integrated mass $\tilde{m}(r)$ contained within a coordinate radius $r$ inside the star. On the other hand, the stellar structure with Schwarzschild geometry is well known as the TOV equation which relates  density and  pressure in a unique way\cite{tolman,ov},
\be
  \frac{\d P(r)}{\d r} = -\frac{1}{r^2} \left[ \rho(r) + P(r) \right] 
  \left[ 4 \pi G \, P(r) \, r^3 + G \, m(r) \right]
  \left[ 1 - \frac{2 G m(r)}{r}  \right]^{-1} \, .
  \label{eq:tov}
\ee

Substituting Eqs. (\ref{eq:e}) and (\ref{eq:p}) into Eq. (\ref{eq:tov}) provides a direct relationship between density and pressure. In another word, one can obtain a contour of the equation of state of gluonic plasma in term of distribution function $\phi(r)$ and temperature $T$.

\section{Summary}

The phase transition between stable hadronic state and highly energized QGP state is briefly discussed using recently developed fluid QCD lagrangian. It is argued that inside the hadronic state the gluons behave as point particles and the properties are determined by its polarization vectors. However, in the hot QGP  state, the gluon should behave as fluid particles and characterized by its relativistic velocities.

Then, the energy momentum tensor for the fluid particle is investigated. In particular, a detailed derivation has been worked out in the case of Schwarzschild space-time that is relevant for stellar structure of unmatured stars composed of hot QGP. In principle one can obtain a kind of equation of state of gluonic plasma inside the star through the TOV equation. This approach would enable us to describe the stellar structure without assuming a particular relation between pressure and density.

Further works, especially on the numerical analysis, are still in progress and will be reported elsewhere.

\section*{Acknowledgments}
The authors thank A. O. Latief, C. S. Nugroho and M. K. Nurdin for fruitful discussion during the final stage of this paper. This work is funded by Riset Kompetitif LIPI in fiscal year 2010 under Contract no.  11.04/SK/KPPI/II/2010.

\bibliographystyle{ws-procs975x65}
\bibliography{qgp}

\end{document}